\documentclass[
  reprint,
  aps,
  pre,
  amsmath,
  amssymb
]{revtex4-2}

\usepackage{graphicx}
\usepackage{booktabs}

\begin{document}

\title{Emergence of polymorphism in stochastic evolutionary games}

\author{Bill Nunn$^\star$, Marcel Ortgiese$^\star$, Tim Rogers$^\star$}
\affiliation{$^\star$Department of Mathematical Sciences, University of Bath, Bath, BA2 7AY, United Kingdom}

\email{Bill Nunn: wn256@bath.ac.uk}


\begin{abstract}

Deterministic evolutionary game theory makes no distinction between a monomorphic population of individuals all of whom share a mixed evolutionarily stable strategy and a polymorphic population of players of pure strategies present in a ratio that reproduces the mixed strategy on average. The so-called trembling hand hypothesis posits that in finite populations demographic noise selects for monomorphism, however, simulation studies have found contradictory results in some situations. Here we resolve this discrepancy by conducting a theoretical analysis of the paradigmatic Hawk-Dove game using timescale separation. We characterise the emergence of polymorphism driven by stochastic effects, finding long-lasting polymorphic states in certain conditions.

\end{abstract}

\maketitle

\section{Extended Replicators}\label{sec:int}

The Hawk-Dove game models conflict within a population of animals for a limited resource \cite{Smith1973}. Animals play the game in pairs, each picking either a `Hawk' action or a `Dove' action. When a Hawk action faces a Dove action, the Hawk chases away the Dove and claims a resource with payoff $V>0$, the Dove receives payoff $0$. When a Dove faces another Dove the resource is peaceably split, and each receives payoff $V/2>0$. When a Hawk faces another Hawk a fight invariably occurs, and each receive payoff $(V-C)/2$. Half the resource $V/2$ is gained, but the cost of fighting $C/2$ is also incurred. If $C > V$ and one knows the other player will pick a Hawk action, it is better to avoid fight, and thus neither pure strategy is dominant in the game theoretic sense. We keep track of these four possible outcomes of each combination of actions in the following payoff matrix
\begin{equation}\label{eq:hd}
    M := \begin{pmatrix}
    \frac{V-C}{2} & V \\
    0 & \frac{V}{2}
    \end{pmatrix}.
\end{equation}

Classical evolutionary game theory postulates that the per capita increase in Hawk actions should be proportional to the expected payoff of Hawk \cite{Hofbauer1998}. Keeping track of the relative frequency for which Hawk is picked as $x_H$, the only ODE which satisfies the classical postulate is the replicator equation
\begin{equation}\label{eq:rep}
\begin{aligned}
    \frac{dx_H}{dt} &= x_H\left( \boldsymbol{e}_H^T M \boldsymbol{x} - \boldsymbol{x}^T M \boldsymbol{x}\right), \\
    &=: x_H \left(f_H - \phi \right)
\end{aligned}
\end{equation}

\begin{figure}
    \centering
    \includegraphics[width=0.72\linewidth]{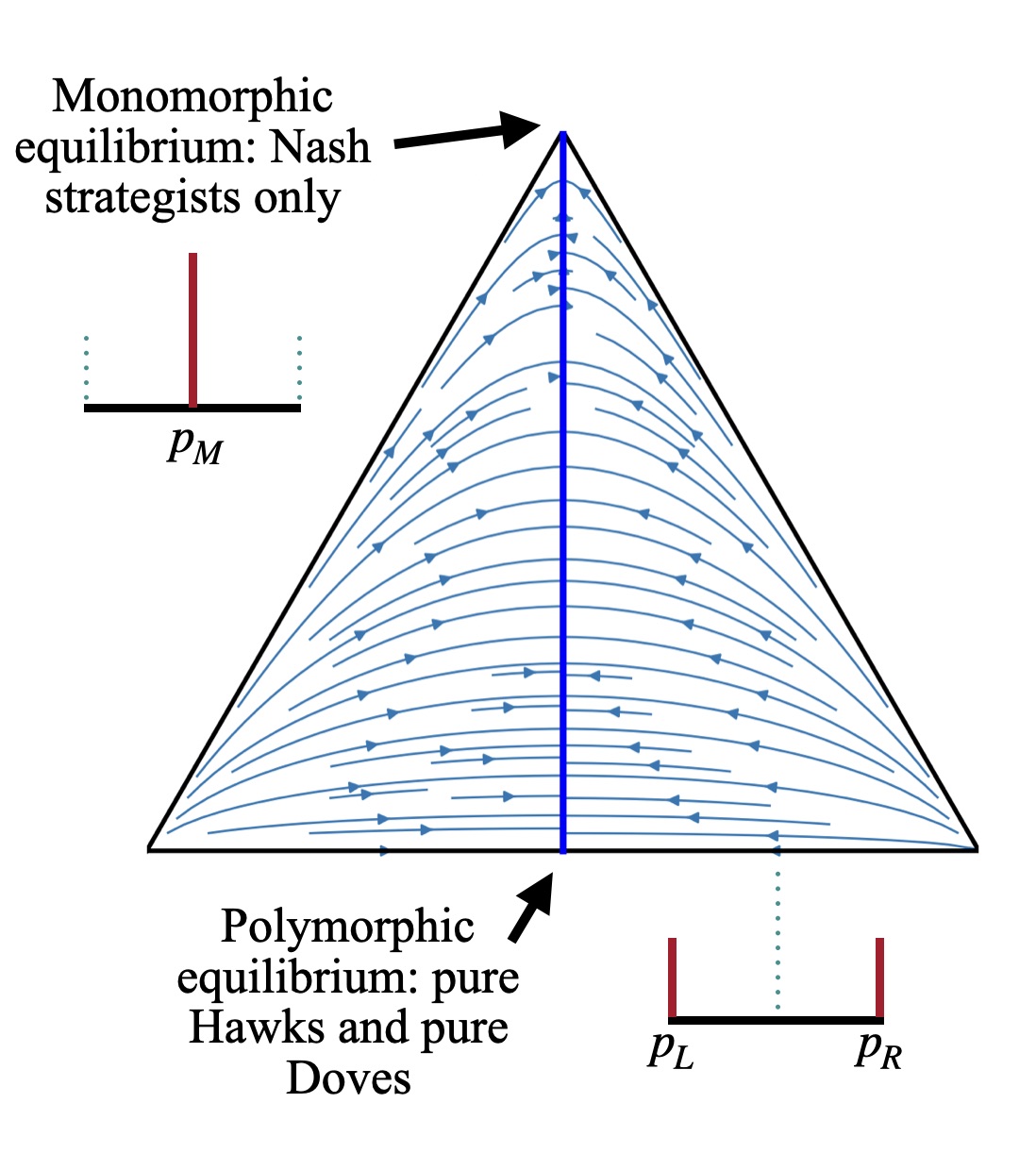}
    \caption{Flow field for the extended Hawk-Dove-Nash replicator system with $2V=C$. The proportions of three types are tracked: pure Dove, pure Hawk, and Nash strategists. The globally attracting centre manifold is plotted in royal blue.}
    \label{fig:hdn}
\end{figure}
In equation \eqref{eq:rep} the column vectors $\boldsymbol{x} := (x_H, 1-x_H)$ give the relative frequency at which Hawk and Dove actions are picked by the population of animals, and $\boldsymbol{e}_H := (1,0)$. Defining the expected payoff of Hawk $\boldsymbol{e}_H^T M \boldsymbol{x}$ as the fitness $f_H$ we see that the replicator has an intuitive form: the per capita increase in the Hawk action is proportional to the difference between the fitness of Hawk and the weighted average of fitness $\phi$ of Hawk and Dove.

The replicator equation \eqref{eq:rep} for Hawk-Dove has a single stable steady state at $x_H = V/C$. Whenever both actions are present in the initial population the relative frequency of Hawk evolves toward and settles at this steady state. The mixed strategy $(V/C, 1-V/C)$ is also the unique internal Nash equilibrium of the game \cite{Broom2022}.

Classical replicators say little about what strategies individual animals play. There are many ways a relative frequency $V/C$ of Hawk can be realised in a population. Every animal in the population could play the mixed strategy $(V/C, 1-V/C)$, or a proportion $V/C$ of the animals could always pick Hawk, and the rest always pick Dove for example. In the former case all animals have an identical strategy and the population is called monomorphic, in the latter case animals can take one of two strategies and the population polymorphic. This subtlety has generally been quite neglected in previous studies of replicators, but is highlighted clearly in \cite{Bergstrom1998}. 

As our work is concerned with the strategies individual animals in the population actually play, we shift perspective from the classical replicator slightly and consider the proportion of animals in a population which play one of three mixed Hawk-Dove strategies. Let the mixed strategies be called $L$ (left), $M$ (middle), and $R$ (right) and have the probability of playing Hawk be $p_L <p_M < p_R$ respectively. Additionally we assume $p_M \leq V/C$, so that $L$ and $M$ are both left of the Nash when the corresponding probabilities are plotted on the line. The extended replicator system \cite{Broom2022}, which tracks the proportion of the population which play each strategy is then given by
\begin{equation} \label{eq:extrep}
\begin{aligned}
\frac{dx_L}{dt} &= x_L \left(\boldsymbol{e}_L^T A \boldsymbol{x} - \boldsymbol{x}^T A \boldsymbol{x}\right),\\
\frac{dx_M}{dt} &= x_M\left( \boldsymbol{e}_M^T A \boldsymbol{x} - \boldsymbol{x}^T A \boldsymbol{x}\right),\\
\frac{dx_R}{dt} &= x_R\left(\boldsymbol{e}_R^T A \boldsymbol{x} - \boldsymbol{x}^T A \boldsymbol{x}\right).
\end{aligned}
\end{equation}

In system \eqref{eq:extrep} the standard unit vectors are ordered such that $\boldsymbol{e}_L := (1,0,0)$, $\boldsymbol{e}_M := (0,1,0)$, $\boldsymbol{e}_R := (0,0,1)$. Matrix entry $A_{i,j}$ gives the expected payoff of an $i$ strategist against a $j$ strategist. We have the constraint $x_L + x_M + x_R = 1$ and thus system \eqref{eq:extrep} is two dimensional. Like standard replicator dynamics, the extended replicator terms
\begin{equation}\label{eq:fit}
    f_i := \boldsymbol{e}_i^T A \boldsymbol{x} \quad \text{for } i \in \{ L, M, R\},
\end{equation}
are the fitness of each strategy when the population is in state $\boldsymbol{x}$. It immediately follows then that $\phi := \boldsymbol{x}^T A \boldsymbol{x}$ is (still) the average fitness.

The flow field of the extended replicator \eqref{eq:extrep} in the case the $L$ strategy is pure Dove, the $M$ strategy is the mixed Nash strategy, and the $R$ strategy is pure Hawk is shown in Figure \ref{fig:hdn}. We refer to the system composed of these three specific strategies as the Hawk-Dove-Nash system. The Hawk-Dove-Nash system \eqref{eq:extrep} quickly equilibrates to a point on the centre manifold \cite{Glendinning1994} of steady states, plotted in royal blue. Notice that the curvature of the flow field will causes the proportion of Nash strategists to increase if starting at a point away from the centre manifold. Let us imagine we have equal starting proportions of the three strategists, that is we start in the middle of the blue centre manifold in Figure \ref{fig:hdn}. Upon inclusion of noise induced by a finite population $K$, the curvature of the flow field tempts us to postulate that the state progresses toward the monomorphic endpoint where all animals use the Nash strategy. Such reasoning, as well as appeals to the `trembling hand' \cite{Broom2022, Hofbauer1998}, has been used to assert that as the population of animals $K$ becomes large the probability that the the monomorphic endpoint is reached tends to unity. The simulated behaviour in \cite{Bergstrom1998} proves different however: the all Nash state is favoured, but the probability it is reached remains bounded away from one for arbitrarily large populations $K$.

Infinite population models like replicators and appeals to trembling hands are very classical, and significant analytic progress has been made working with (stochastic) finite population models (and their SDE approximations) directly \cite{Traulsen2009}. The essential theme of this body of work is that in finite population models demographic noise can induce interesting phenomena which are not predicted by the limiting models, including reversal of selection \cite{Constable2016} and spontaneous speciation \cite{Rogers2012}. Our work continues in this tradition: we formulate a finite population model of extended Hawk-Dove, find a corresponding SDE, and calculate the fixation probabilities for large $K$ in Hawk-Dove-Nash using a timescale separation. The apparent discrepancy between the classical prediction and simulated behaviour in \cite{Bergstrom1998} is fully resolved analytically.

\section{Individual based model and corresponding SDE}\label{sec:ibm}

\begin{figure}
    \centering
    \includegraphics[width=0.72\linewidth]{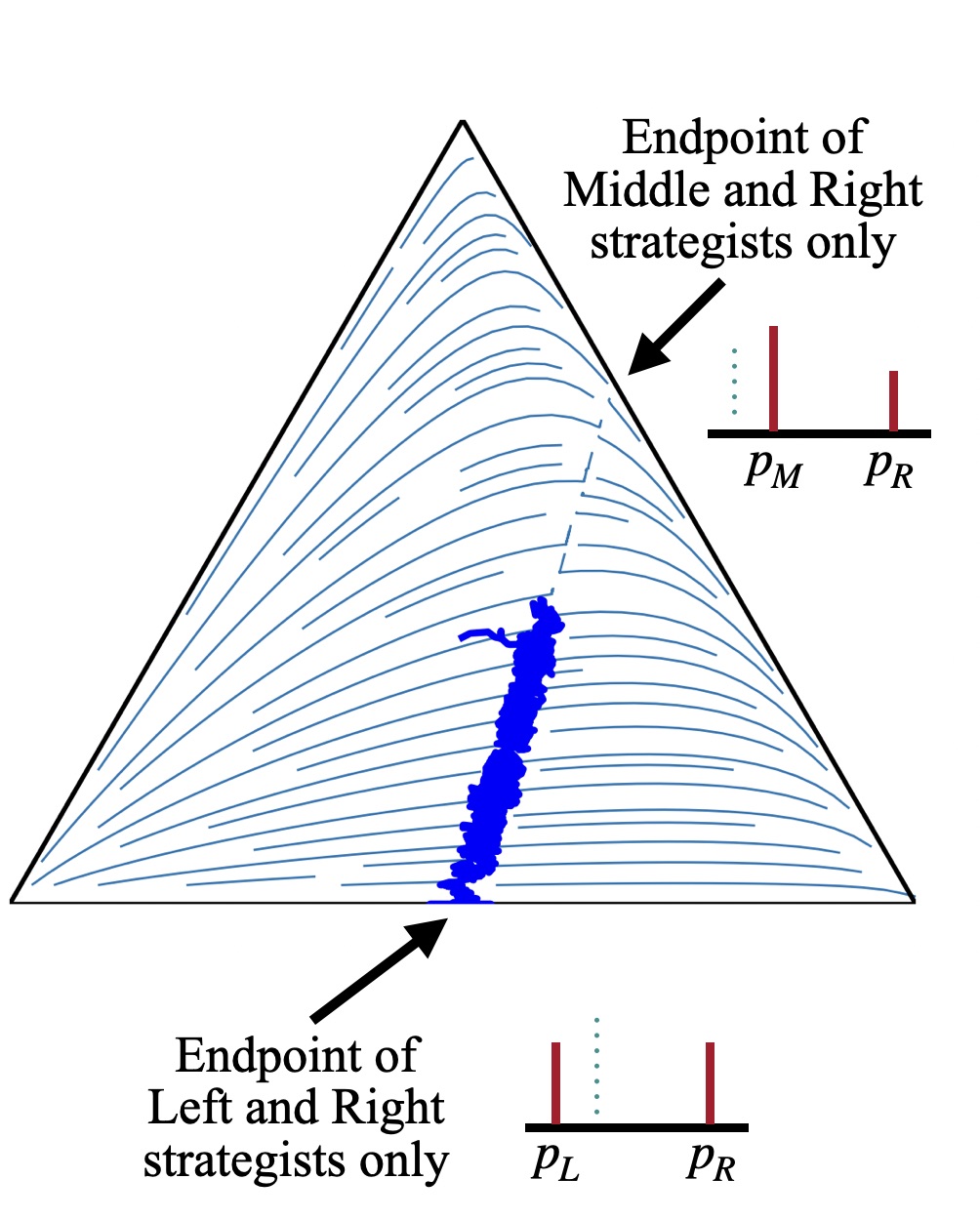}
    \caption{A sample path of SDE \eqref{eq:sde} in blue, fixating at a polymorphic equilibrium composed of $L$ and $R$ strategists. We see the system remains close to the centre manifold. Parameters for this simulation are $C=2V=1/2$, $K=2000$, $p_L = 0.1$, $p_M = 0.3$, $p_R=0.9$. The sample path was started with all strategies in equal proportion, the centre of the ternary plot.}
    \label{fig:sde}
\end{figure}

We formulate a finite population model of Hawk-Dove in the style of Traulsen et al.~\cite{Traulsen2006, Traulsen2012}, with animals able to take one of three arbitrary mixed strategies. Again, following Traulsen et al.’s example we approximate the finite population model with SDEs, using methodology of Kurtz \cite{Kurtz1978, Baxendale2010} to derive said SDEs, making for an alternative to the system size expansion approach popularised by van Kampen \cite{Kampen2007}. The deterministic drift of the obtained SDE system is shown to be extended replicator dynamics.

To begin, we assume a fixed finite population of size $K$ and that pairs of individuals are chosen uniformly at random to interact. In other words, any pair of individuals can interact and the population is well mixed. Here the time between interactions is given by an exponential random variable with rate $K$,
so that in expectation  each individual has a constant order of interactions per unit time.

Suppose that an $i$ strategist and a $j$ strategist are chosen to interact. The fitness \eqref{eq:fit} of each strategist is compared and the $i$ strategist replaces the $j$ strategist with probability 
\begin{equation}\label{eq:pair}
\gamma(f_i, f_j) := \frac{1}{2}\left(1 + \frac{1}{2C}(f_i - f_j)\right).
\end{equation}
The denominator $2C$ ensures that the probability $\gamma$ sits between zero and one. As one should expect, this choice of $\gamma$ satisfies $\gamma(z,y) = 1 - \gamma(y,z)$. Since interactions occur at rate $K$ we can specify the individual based model in terms of the following reactions
\begin{equation}
    i + j \to 2i \quad \text{at rate} \quad \frac{n_i n_j \cdot \gamma(f_i, f_j)}{K},
\end{equation}
for all combinations of $i, j \in \{L,M,R\}$, where $n_i$ is the number of $i$ strategists. 

Following Kurtz \cite{Kurtz1978}, we now define a rescaled jumping rate $\Psi(\boldsymbol{x} ; \boldsymbol{r})$ in terms of the proportions of each strategy $x_i = n_i / K$. When a $L$ strategist replaces a $M$ strategist the jump will be $(+1,-1,0)$ for example, with other jumps defined accordingly. We find a rescaled jumping rate kernel $\Psi(\boldsymbol{x}; \boldsymbol{r})$ for all possible jumps $\boldsymbol{r}$. Defining the value of the kernel $\Psi$ for jump $(+1,-1,0)$ as
\begin{equation}
    \Psi(\boldsymbol{x}; (+1,-1,0)) := x_L x_M \cdot \gamma(f_L, f_M),
\end{equation}
and the other jumping rates accordingly, yields the following Markov jump process for the evolution of the state of the system
\begin{equation}\label{eq:mjp}
    \boldsymbol{x}(t) = \boldsymbol{x}(0) + \frac{1}{K}\sum_{\boldsymbol{r}} \boldsymbol{r}N_{\boldsymbol{r}} \left( \int_0^t K \Psi \left( \boldsymbol{x}(s) ; \boldsymbol{r} \right) \,ds \right).
\end{equation}
Here the $N_{\boldsymbol{r}}$ are independent unit rate Poisson counting processes for each type of jump. Equation \eqref{eq:mjp} is intuitively clear: the proportion of individuals in each strategy is given by the number of each type of jump multiplied by the rescaled effect of each type of jump.

The essential approximation Kurtz \cite{Kurtz1978} uses to pass from jump processes to diffusions is that $(N(tK)-tK)/\sqrt{K} \approx B(t)$ for $K$ large and $B$ a Brownian motion.
The integral on the right hand side of \eqref{eq:mjp} is linear in $K$, and thus when $K$ is large this diffusive approximation is appropriate. Invoking the approximation and working through the random time change yields an SDE of the form
\begin{equation}\label{eq:sde}
d \boldsymbol{x} = F(\boldsymbol{x}) dt + \frac{1}{\sqrt{K}} D(\boldsymbol{x}) dW, 
\end{equation}
Where $W$ is a three dimensional Brownian motion, matrix $D$ satisfies $D^TD = B$, and we have 
\begin{align}
F_i(\boldsymbol{x}) &= \sum_{\boldsymbol{r}}\Psi \left( \boldsymbol{x}(s) ; \boldsymbol{r} \right)r_i,\\
B_{ij}(\boldsymbol{x}) &= \sum_{\boldsymbol{r}} \Psi \left( \boldsymbol{x}(s) ; \boldsymbol{r} \right)r_i r_j.
\end{align}
See Baxendale and Greenwood \cite{Baxendale2010} for a full derivation of these relations. 

For our particular individual based model we find the drift for $x_L$ is given by
\begin{align*}
    F_L(\boldsymbol{x}) &= x_L \left[ x_M(2\gamma(f_L,f_M) - 1) + x_R(2\gamma(f_L,f_R) - 1)\right]\\
    &= \frac{1}{2C} x_L\left[x_M(f_L-f_M) + x_R(f_L - f_R) \right]\\
    &= \frac{1}{2C} x_L\left[f_L(x_L + x_M + x_R) - \boldsymbol{x}^T \boldsymbol{f} \right]\\
    &= \frac{1}{2C} x_L\left[f_L - \phi \right].
\end{align*}
The drift of the SDE \eqref{eq:sde} is thus a linear time rescaling of the extended replicator system \eqref{eq:extrep}. This same result was obtained by Traulsen \cite{Traulsen2006} in the non-extended case, but by using a system size expansion instead.

Matrix $B$ is given by
\begin{equation*}\label{eq:noi}
B(\boldsymbol{x}) = \begin{pmatrix}
x_L(1-x_L) & -x_L x_M & -x_L x_R\\
-x_L x_M & x_M(1-x_M) & -x_M x_R\\
-x_L x_R & -x_M x_R & x_R(1-x_R)
\end{pmatrix},
\end{equation*}
in our case. We remark that this matrix is invariant for any choice of $\gamma$ which satisfies the relation $\gamma(y,z) = 1 - \gamma(z,y)$.

The expected payoff of all strategies against the Nash strategy $(V/C, 1-V/C)$ is identical, and thus if the population average strategy is Nash we are at a stationary point the drift of \eqref{eq:sde} (i.e. the extended replicator system). The centre manifold $\Lambda$ of the extended replicator system is thus given by all points $\boldsymbol{y}$ which satisfy
\begin{equation}\label{eq:stat}
    y_Lp_L + y_Mp_M + y_Rp_R = 
    V/C.
\end{equation}
Using \eqref{eq:stat} and the constraint $y_R = 1-y_L-y_M$ we see that the centre manifold $\Lambda$ is the the straight line
\begin{equation}\label{eq:line}
    y_M = \lambda(y_L) = \frac{p_RC - V}{p_R-p_M} - \frac{p_R-p_L}{p_R-p_M}y_L.
\end{equation}

Figure \ref{fig:sde} plots a sample path of the SDE \eqref{eq:sde}. We see the state does not deviate far from the centre manifold. When the trajectory reaches either endpoint of the manifold it cannot return to the interior. Assuming that $M$ is left of the Nash, the two endpoints of the manifold in Figure \ref{fig:sde} correspond to elimination of the $L$ strategy, or elimination of the $M$ strategy respectively. Either way, the $R$ strategy is not initially eliminated (with very high probability) and the system enters a metastable state of two strategies, with one on each side of the Nash.

\section{Reduced SDE for three strategies}\label{sec:red}

When three mixed strategies are present in the population we have seen that the deterministic drift of \eqref{eq:sde} (i.e. the extended replicator system) has a globally attracting centre manifold of steady states $\Lambda$. If the population $K$ is large then the noise (which scales like $1/\sqrt{K}$) is small, and thus the dynamics of the SDE system progress close to the manifold. Indeed, Katzenberger \cite{Katzenberger1991} proves as $K$ is taken to infinity trajectories of such dynamics converge to the stochastic variable with law given by an SDE on $\Lambda$. We call this SDE the \emph{reduced} SDE. Parsons and Rogers \cite{Parsons2017} give a toolbox of methods for obtaining the reduced system explicitly. We add a tool to the toolbox by exploiting a nice property of replicator dynamics: we can find a function whose level sets are the flow field. Reduction to dynamics on lower dimensional manifolds is an increasingly popular technique in evolutionary game theory, see \cite{Parsons2018, Popovic2020, Bhat2025, Oliveira2026} for recent examples.

Intuitively, the trajectories of SDE \eqref{eq:sde} for large $K$ can be broken into two stages. In the first stage a stochastic bump takes the state away from the manifold. In the second stage the system quickly re-equilibrates, following the deterministic flow line back to the manifold. It is consideration of the projection by the flow lines to $\Lambda$ which allows us to find expressions for the drift and diffusion of the reduced SDE. Let $\boldsymbol{\pi}(x_L, x_M) = (\pi_L(x_L, x_M), \pi_M(x_L,x_M))$ be the projection onto $\Lambda$ by the flow lines of the extended replicator. We let $\boldsymbol{y} = (y_L, y_M)$ be a point on $\Lambda$. Defining the functions 
\begin{align}\label{eq:ps}
P_{ij}(\boldsymbol{y}) &:= \frac{\partial \pi_i(\boldsymbol{x})}{\partial x_j}\Big\vert_{\boldsymbol{x}=\boldsymbol{y}}, \\
Q_{ijk}(\boldsymbol{y}) &:= \frac{\partial^2 \pi_i(\boldsymbol{x})}{\partial x_j \partial x_k} \Big\vert_{\boldsymbol{x}=\boldsymbol{y}},\label{eq:qs}
\end{align}
will allow us to express the reduced SDE.

The reduced SDE is given by
\begin{widetext}
\begin{equation}\label{eq:katz}
dy_i = \frac{1}{2K} \sum_{s,j,k\in \{L,M\}} D_{js}(\boldsymbol{y})D_{ks}(\boldsymbol{y})Q_{ijk}(\boldsymbol{x})dt + \frac{1}{\sqrt{K}} \sum_{s,j\in \{L,M\}} P_{ij}(\boldsymbol{y}) D_{js}(\boldsymbol{y}) dW^s_t \quad \quad \forall i \in \{L,M\},
\end{equation}
\end{widetext}
where $(W^L,W^M)$ is a two dimensional Brownian motion.
Katzenberger proves convergence to the reduced SDE \eqref{eq:katz} but does not discuss how to calculate partial derivatives of the projection $\boldsymbol{\pi}$. Calculating these derivatives is not trivial, and no general method appears in the literature until \cite{Parsons2017}. We use a different method to calculate the derivatives, which works in the (rather special) case of replicator dynamics.

Parsons and Rogers \cite{Parsons2017} assume that the flow lines cannot be obtained by simple means, and thus second order approximation of the flow lines is necessary. We are more fortunate: the flow lines of the extended replicator system \eqref{eq:extrep} can be expressed as level sets of a closed form function. More precisely, the function
\begin{widetext}
\begin{equation}
    g(\boldsymbol{x}) := \frac{(p_R-p_M)\left[(p_R-p_M)\log (x_L) + (p_M-p_L) \log(1-x_L-x_M) - (p_R-p_L)\log(x_M)\right]}{(p_R-p_L)(3p_R-2p_M-p_L)}.
\end{equation}
\end{widetext}
is a constant of motion of the ODE
\begin{equation}
    \frac{dx_M}{dx_L} = \frac{F_M(\boldsymbol{x})}{F_L(\boldsymbol{x})}.
\end{equation}

Recall that the centre manifold $\Lambda$ can be parametrised by the first coordinate of the system, as in \eqref{eq:line}. If we know the value of the first coordinate is $y_L$ then the second coordinate is uniquely determined we will denote it by $\lambda(y_L)$. Since $g$ is constant along flow lines, and $\boldsymbol{\pi}$ projects to $\Lambda$ along said flow lines, we have $\pi_M = \lambda(\pi_L)$ and
\begin{equation}\label{eq:key}
    g(\boldsymbol{x}) = g(\pi_L(\boldsymbol{x}), \lambda(\pi_L(\boldsymbol{x}))).
\end{equation}
Differentiating \eqref{eq:key} with respect to $x_L$ yields
\begin{widetext}
\begin{equation}\label{eq:diff}
        \frac{\partial g(\boldsymbol{x})}{\partial x_L} = \frac{\partial \pi_L(\boldsymbol{x})}{\partial x_L} g^{(1,0)}(\pi_L(\boldsymbol{x}), \lambda(\pi_L(\boldsymbol{x})))\\ + \frac{\partial \pi_L(\boldsymbol{x})}{\partial x_L} \lambda'(\pi_L(\boldsymbol{x})) g^{(0,1)}(\pi_L(\boldsymbol{x}), \lambda(\pi_L(\boldsymbol{x}))).
\end{equation}

\end{widetext}
Here $g^{(i,j)}(\cdot,\cdot)$ denotes the $i$th partial derivative of $g$ with respect to the first argument, and $j$th partial derivative with respect to the second argument, evaluated at $(\cdot, \cdot)$. Recall we only require the partial derivatives of the projection $\boldsymbol{\pi}$ evaluated at points $\boldsymbol{y}$ on the manifold. At such points however we have $\boldsymbol{y} = (\pi_L(\boldsymbol{y}), \lambda(\pi_L(\boldsymbol{y})))$ and hence
rearranging \eqref{eq:diff} yields a closed form expression for the partial derivative of $\pi_L$ with respect to $x_L$ evaluated at $\boldsymbol{x} = \boldsymbol{y}$. 

Similar expressions can be found for the remaining partial derivatives of the projection \eqref{eq:ps}, \eqref{eq:qs} using the same strategy: take derivatives of \eqref{eq:key} and rearrange. The expressions for the derivatives are unwieldy, but all that matters for our purposes is that they can be expressed in closed form. The drift coefficient of $y_L$ in the reduced SDE \eqref{eq:katz} can then also be found in closed form, namely
\begin{widetext}
\begin{equation}\label{eq:drift}
    \mu(y_L):=\frac{1}{2 K}\frac{C^2 (p_M - p_L) (p_R - p_M) y_L (V - C p_M (1-y_L) - C p_L y_L) (V - C p_R (1-y_L) - C p_L y_L)}{((V - C p_M) (V - C p_R) - C^2 (p_M - p_L) (p_R - p_L) y_L)^2}.
\end{equation}
\end{widetext}

Equation \eqref{eq:drift} is a slightly strange beast. The first thing to note is that $\mu(y_L)$ is negative throughout its domain and thus fixation of $M$ strategists is more likely than neutral. Less intuitively, for fixed $K, p_L, p_M, p_R$ the drift $\mu(y_L)$ only depends on the ratio $V/C$. A moment's thought reveals sense however: our reduced model only depends upon the projection to the centre manifold along the flow lines, but the flow lines only depend on $V/C$. In other words, the strength of the drift of the unreduced SDE is of no consequence, so long as population $K$ is large enough.

The noise term of the reduced SDE \eqref{eq:katz} is expressed in terms of the sum of two independent Brownian motions, and for convenience we re-express it in terms of a single Brownian motion with the same variance. Let us denote the diffusion coefficient for the reduced SDE \eqref{eq:katz} as $\sigma(y_L)$. Just like the drift, we can find a closed (albeit very unwieldy) form expression for $\sigma(y_L)$.

\section{Hawk-Dove-Nash}\label{sec:res}

Hawk-Dove games with any three mixed strategies have been reduced to a one dimensional SDE. Let us enjoy the fruits of our labour and investigate Hawk-Dove dynamics when the mixed Nash strategy is included. Throughout this section we let $2V = C$. This choice is not restrictive, it just fixes the Nash strategy to be $(1/2,1/2)$.

Recall in Hawk-Dove-Nash the population is composed of pure Hawks, pure Doves and (mixed) Nash strategists. In this case the left $L$ strategy is pure Dove, the middle $M$ is Nash, and the right $R$ pure Hawk. The reduced SDE for $y_D$ (the proportion of the population which are pure Doves) takes values in $[0,1/2]$. If the reduced SDE reaches the $0$ endpoint then the Nash strategists fixates, instead if the reduced SDE reaches $1/2$ endpoint then the Nash strategy is eliminated, leaving a polymorphic mixture of pure strategists. We calculate the probability of each outcome analytically.

\begin{figure}
    \centering
    \includegraphics[width=1.0\linewidth]{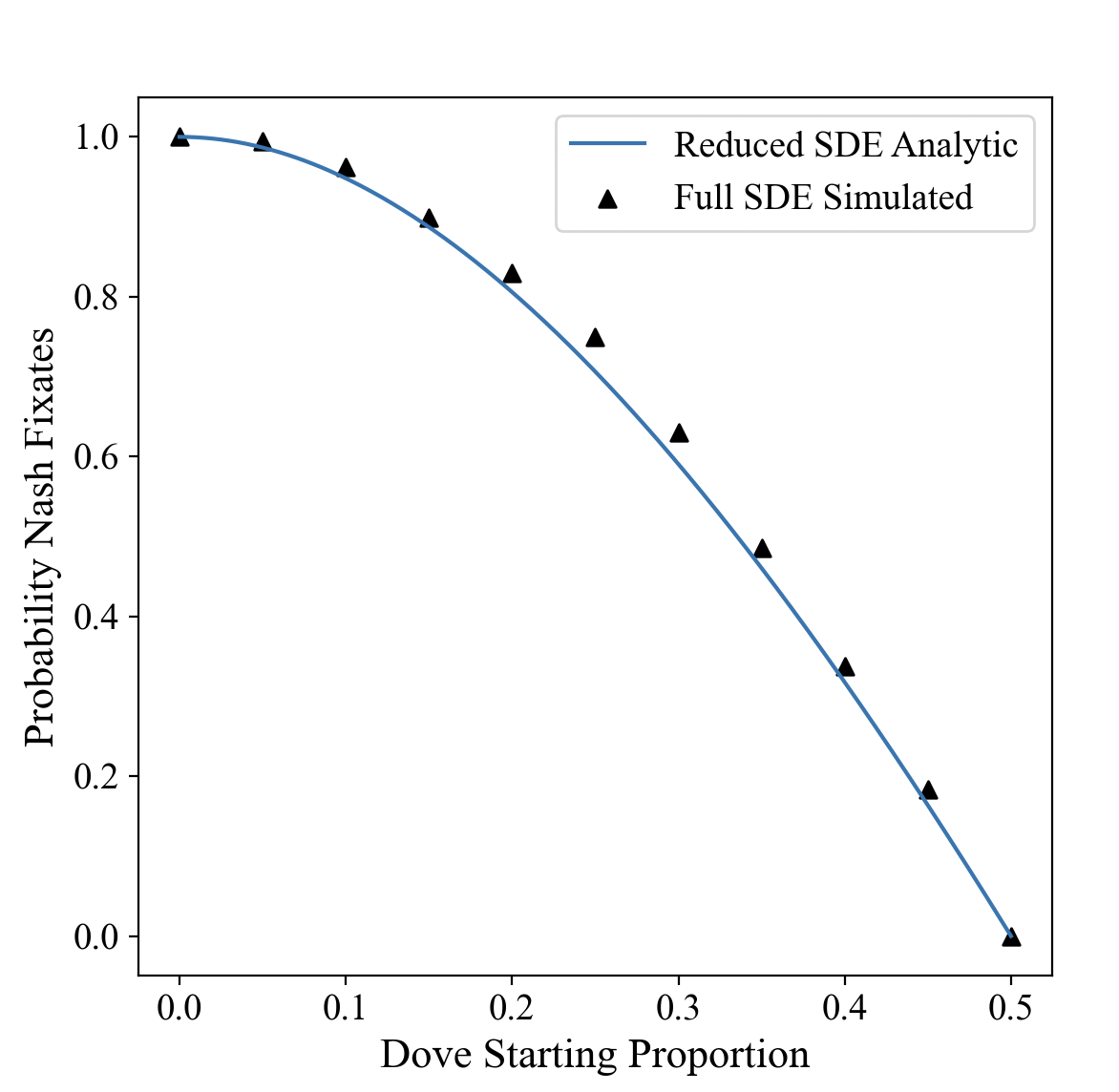}
    \caption{Comparison of empirical fixation probabilities of the unreduced SDE system ($K=2000$, $2V=C=1$, one thousand runs for each starting proportion) with the analytic estimate \eqref{eq:hit} derived from the reduced SDE.}
    \label{fig:fix}
\end{figure}

Supposing the one dimensional reduced SDE has drift $\mu(y_D)$ and diffusion $\sigma(y_D)$ we find the scale function $S(y)$ (see e.g.~Etheridge \cite{Etheridge2011}) which satisfies
\begin{equation}
    \mathbb{P}[T_0 < T_{1/2} | y_D(0) = y] = \frac{S(1/2) - S(y)}{S(1/2) - S(0)},
\end{equation}
where $T_0$ is the (stopping) time that the SDE hits endpoint $0$, and $T_{1/2}$ the endpoint $1/2$. The scale function can be expressed as
\begin{equation}\label{eq:scale}
    S(y) = \int_0^y \int^{\frac{1}{2}}_x \exp \left( \frac{2 \mu(z)}{\sigma^2(z)} \right) dz dx.
\end{equation}
Quite remarkably given the unwieldiness of our expressions for $\mu$ and $\sigma$, for the case of Hawk-Dove-Nash we have $2\mu(z) / \sigma^2(z) = 1-1/z$. Note immediately therefore that fixation of the reduced SDE has no dependence on the population size $K$. Since the reduced SDE is a good approximation for large $K$, we see that the fixation of the individual based model plateaus as $K$ is increased. Evaluating the integral \eqref{eq:scale} for Hawk-Dove-Nash yields $S(y) = \Gamma(2,y)$ where $\Gamma$ denotes the incomplete Gamma function
\begin{equation}
    \Gamma(s,z) := \int^z_0 t^{s-1} e^{-t} dt.
\end{equation}
We thus have for Hawk-Dove-Nash that
\begin{equation}\label{eq:hit}
\mathbb{P}[T_0 < T_{1/2} | y_D(0) = y] = \frac{\Gamma(2,1/2)-\Gamma(2,y)}{\Gamma(2,1/2)}.
\end{equation}
This fixation probability is plotted against simulations of the full SDE system in Figure \ref{fig:fix}. We see agreement between the analytic estimate \eqref{eq:hit} and simulations of the full SDE for $K=2000$. In particular, for a fixed starting proportion of pure Hawk, pure Doves and Nash strategists, the fixation probability of the Nash is bounded away from one. This fully resolves the observations of \cite{Bergstrom1998} analytically. 

It is important to observe from Figure \ref{fig:fix} that Nash strategists can invade the polymorphic mixture, with a single Nash mutant fixating with probability which scales as $1/K$. The probability that a mutant polymorphic mixture fixates in an initial population of Nash strategists is vanishingly small however: the gradient of \eqref{eq:hit} at $y=0$ is zero. In the class of large $K$ models where only pure Hawk, pure Dove, and Nash strategists are possible, we would thus expect the Nash strategists to dominate. The singularity in $2\mu(z) / \sigma^2(z) = 1 - 1/z$ at $z=0$ induces the zero gradient in the curve in Figure \ref{fig:fix}.

\section{Comparison with Adaptive Dynamics}

Adaptive dynamics \cite{Dieckmann1996} deals with traits animals exhibit more directly than replicators do. Indeed, qualitative emergence of trait polymorphisms at evolutionary branching points has been one of the core successes of adaptive dynamics theory \cite{Brannstorm2013}. Continuous versions of Hawk-Dove have been considered by the adaptive dynamics community \cite{Doebeli2004}, where a proxy for the probability of playing Hawk $p \in [0,1]$ is the trait under consideration. In the continuous setting either monomorphic or polymorphic populations can arise depending on the particular structure of the fitness function. The case with closest parallels to the extended replicator, in which fitness is exactly the expected payoff of a given strategist in the population, cannot be dealt with via adaptive dynamics \cite{Dieckmann2006}. For this special case Dieckmann and Metz \cite{Dieckmann2006} show that the all Nash state is the evolutionary endpoint of adaptive dynamics under a fluctuating payoff matrix, or when different strategists have differing interaction rates. We have considered the replicator case under the effect of a finite population instead. See \cite{Wakano2013} for additional discussion of evolutionary branching in a finite populations.

As laid down by Dieckmann and Law \cite{Dieckmann1996}, adaptive dynamics hinges upon the assumption of mutual exclusion. That is, there is a time-scale separation between mutant traits arising and the time taken for a single trait to fixate. In Hawk-Dove however, a population of pure Hawk strategists and pure Dove strategist is metastable, with the time to fixation exponential in population size. We therefore cannot expect mutual exclusion assumption to hold for any sub-exponential choice of mutation timescale. Extending adaptive dynamics to account for metastable states requires careful thought \cite{Champagnat2010, Parsons2018, Sireci2024}. Champagnat and Meleard \cite{Champagnat2010} deal with metastable states in their work on emergence of polymorphic populations under local competition kernels in the trait space. The continuation of our work shall examine the scenario in which competition between traits is mediated by a game instead.

Under a careful choice of mutation timescale, we find that at most three strategies are likely to simultaneously exist, with fixation to (meta)stable states of at most two strategies occurring before new mutants arise. Since we can find probabilities that each of the three strategies is eliminated, the way is paved for stochastic simulations over the evolutionary timescale. Long lasting polymorphic states can occur stochastically under certain conditions, and in a manner distinct from the stochastic evolutionary branching described in \cite{Wakano2013}.

\section{Evolutionary Dynamics}\label{sub:ful}

When two strategies are on the same side of the Nash the time until one of the strategies fixates scales as $K$. Instead, if the two strategies are either side of the Nash the time until one fixates scales as $\exp (K)$: the states comprising of two strategies either side of the Nash are metastable. In the case of three strategies, we have already seen that the first strategy to be eliminated must be from the pair of strategies on the same side of the Nash, resulting in a metastable state of two strategies. Recall that elimination of a strategy corresponds to reaching one an endpoints of the centre manifold in Figure \ref{fig:sde}. The time until the first of the three strategies is eliminated scales as $K$.

Therefore, if the timescale in which mutant strategies arise is super-linear but sub-exponential in $K$ we obtain systems which have at most three strategies simultaneously existing. This choice of mutation timescale is akin to the mutual exclusion assumption of classical adaptive dynamics \cite{Dieckmann1996}. Since our reduced SDE \eqref{eq:katz} well characterises the behaviour of any three strategies we can probe the evolutionary behaviour of systems which have this choice of mutation timescale. In particular, given a (meta)stable state of (at most) two strategies and the strategy of a single mutant individual we can find the probability the mutant strategy fixates by using the closed form expressions of the drift and diffusion of the reduced SDE \eqref{eq:katz} and (numerically) evaluating the integral \eqref{eq:scale}. One could then see how the set of a strategies in the population is likely to evolve via repeated mutation and fixation without having to simulate the full model each time a mutant arises.

To illustrate the unusual array of behaviour present in Hawk-Dove with arbitrary strategies we consider two cases. In the first case we investigate the evolutionary dynamics when there is no population of strategists right of the Nash, and the Nash is a reflecting endpoint for an evolving population left of the Nash. These dynamics change markedly in the second case when we condition on there being a population of strategists right of the Nash.

\subsection{Case One}\label{sub:c1}

Fix $2V=C$ and let the two strategies left of the Nash be given by the resident strategy $(1/2 - p, 1/2 + p)$ and the mutant strategy $(1/2 - p + s, 1/2 + p - s)$. Using the methodology of Section \ref{sec:ibm}, the SDE which tracks the proportion $x$ of the population playing the resident strategy can be shown to be
\begin{equation}\label{eq:2strat}
    dx = -\frac{1}{2}s x(1-x) (p + s - s x) dt + \frac{\sqrt{x(1-x)}}{\sqrt{K}} dW.
\end{equation}
Note that this SDE is already one dimensional and so no reduction is required to calculate hitting probabilities. The drift of \eqref{eq:2strat} is negative if and only $s$ is positive, and thus mutant strategies closer to the Nash are more likely to fixate than neutral. Furthermore, we then find
\begin{equation}\label{eq:int}
    \frac{2\mu(x)}{\sigma^2(x)} = -K s(p + s - s x).
\end{equation}
It is important to note that the above integrand of the scale function is proportional to the population $K$.

Assuming all mutants are forced to remain left of the Nash and all mutants have strategies which are small perturbations centred on the resident strategy, we find the evolutionary dynamics can be approximated by the following Ornstein-Uhlenbeck process with reflecting boundary
\begin{equation}\label{eq:ou}
    dp = - K p dt + dW \quad \text{on } k \in (0,1/2).
\end{equation}
In \eqref{eq:ou} the value $p$ tracks the deviation of the resident strategy away from the Nash over evolutionary time. Observe the dependence of the drift on $K$. In particular, if all mutants are fixed to take strategies on one side of the Nash the expected deviation away from the Nash scales like $1/\sqrt{K}$. For large $K$ then, strategies remain very close to the Nash.

\subsection{Case Two}\label{sub:c2}

The behaviour changes considerably when there is a resident population of strategists on the other side of the Nash. Suppose for simplicity that there is a population of pure Hawks on the right of the Nash. Once again we let the resident strategy left of the Nash be $(1/2 - p, 1/2 + p)$ and the mutant strategy to be $(1/2 - p + s, 1/2 + p - s)$. Using the reduced SDE methodology developed in Section \ref{sec:red} we find that the drift of the corresponding reduced SDE is negative if and only if $s$ is positive. Just like in Subsection \ref{sub:c1} mutants closer to the Nash are still favoured. What changes dramatically however is the strength of the drift relative to the strength of the noise.

We have, for all $p \in (0,1/2)$ and $s<p/2$ that
\begin{equation}
\begin{cases}
0>\frac{2\mu(x)}{\sigma^2(x)}>-b &\quad \text{if } s>0,
 \\
0<\frac{2\mu(x)}{\sigma^2(x)}<b  & \quad \text{if } s<0.
\end{cases}
\end{equation}
Crucially, here $b$ is a small positive constant which does not depend on the population size $K$. By assuming a stronger drift towards the Nash and a weaker drift away from the Nash we obtain a system which remains nearer to the Nash than reality. In particular, letting
\begin{equation}\label{eq:approx}
\begin{cases}
\frac{2\mu(x)}{\sigma^2(x)} = -b & \quad \text{for } s>0,
 \\
\frac{2\mu(x)}{\sigma^2(x)} = b & \quad \text{for } s<0,
\end{cases}
\end{equation}
gives an approximating system with a trait which evolves closer to the Nash than the true system. The evolutionary dynamics for this approximating system \eqref{eq:approx} satisfy
\begin{equation}\label{eq:app}
    dp =  - c dt + dW \quad \text{on } k \in (0,1/2),
\end{equation}
where the value $c$ is another small positive constant which does not depend on $K$. Supposing a reflecting boundary, the average deviation away from the Nash strategy in this case scales like $1/c$ and we can therefore observe much larger deviations away from the Nash than the first case. Conditioning on having some population on the other side of the Nash, we can thus observe emergence of polymorphism driven by noise.

\begin{acknowledgments}
B.N. is supported by a scholarship from the EPSRC Centre for Doctoral Training in Statistical Applied Mathematics at Bath (SAMBa), under the project EP/S022945/1. M.O. is partially supported by EPSRC research grant EP/X040089/1.
\end{acknowledgments}

\bibliography{apssamp}

\end{document}